\documentclass[aps,prd,floatfix,tighten,nofootinbib,preprint,showpacs,a4paper]{revtex4-1}
\usepackage{graphicx}
\usepackage{epic}
\usepackage{eepic}
\usepackage{latexsym}
\usepackage{upgreek}
\usepackage{color}
\usepackage{multirow}
\newcommand{\be}{\begin{equation}}
\newcommand{\ee}{\end{equation}}
\newcommand{\bea}{\begin{eqnarray}}
\newcommand{\eea}{\end{eqnarray}}

\begin{document}

\title{Inside the Schwarzschild-Tangherlini black holes}

\author{Jerzy Matyjasek}
\email[]{jurek@kft.umcs.lublin.pl}
\affiliation{Institute of Physics, Maria Curie-Sk{\l}odowska University\\
pl. Marii Curie-Sk\l odowskiej 1, 20-301 Lublin, Poland}
\author{Pawe{\l} Sadurski}
\affiliation{Institute of Physics, Maria Curie-Sk{\l}odowska University\\
pl. Marii Curie-Sk\l odowskiej 1, 20-301 Lublin, Poland}

\date{\today}

\begin{abstract}
The first-order semiclassical Einstein field equations are solved in the 
interior of the Schwarzschild-Tangherlini black holes. The source term is taken
to be the stress-energy tensor of the quantized massive scalar field
with arbitrary curvature coupling calculated within the framework of the
Schwinger-DeWitt approximation. It is shown that for the minimal coupling
the quantum effects tend to  isotropize the interior of the black hole 
(which can be interpreted as an anisotropic collapsing universe) 
for $D=4$ and $5,$ whereas for $D=6$ and $7$ the spacetime becomes more anisotropic.
Similar behavior is observed for the conformal coupling with the reservation
that for $D=5$  isotropization of the spacetime occurs during (approximately) the first $1/3$
of  the lifetime of the interior universe.
On the other hand, we find that regardless of the dimension,
the quantum perturbations initially strengthen the grow of curvature
and its later behavior depends on the dimension and the coupling.
It is shown that the  Karlhede's scalar can still be used as a useful device
for locating the horizon of the quantum-corrected  black hole, as expected.

\end{abstract}
\maketitle

\section{Introduction}
Although described by the same line element, the classical interior of the 
Schwarzschild-Tangherlini~\cite{tangh} black hole has entirely different 
properties than the region outside the event horizon and can be better 
understood as some sort of the anisotropic and nonstatic 
universe~\cite{novikov1961,brehme1977,doran2008}. This interpretation 
(not mandatory) is  helpful when one is forced to abandon usual, i.e., 
referring to the external world, interpretation of the coordinates,  
metric potentials and so on. In the  $D=4$-dimensional case much work 
have been done in this direction and we have a good understanding
of the geometry and dynamics of the classical interior (See e.g., Refs.~
\cite{DiNunno,Frolov2007,Chri,Ingemar} and the references therein). On the other 
hand, less is known about quantum processes inside black holes and their influence upon 
the background geometry.

In the recent paper~\cite{Matyjasek2015} we have studied influence of the quantized 
fields on the static spacetime of the Schwarzschild-Tangherlini black hole 
using the semi-classical Einstein field equations. Since the  stress-energy 
tensor constructed in that paper functionally depends on the metric,
one has a rare opportunity to analyze and compare the quantum corrections 
to the black hole characteristics (and the geometry itself) calculated for 
various spacetime dimensions. The purpose of this paper, which is a natural continuation 
of Refs.~\cite{Matyjasek2015,HiscockInside1997},  is to extend the study
of  the quantized fields to the interiors of the higher-dimensional black holes. 
It should be noted however, that 
now  there are problems that do not appear in the external region. 
The first one is the problem of the central singularity and its closest vicinity. 
It is evident that the semicalssical Einstein field equations
cannot be trusted there. The second difficulty is to some extend related to the previous
one and may be stated as follows. The effective action of the quantized fields 
for $r_{+} \geq 0$  ($r_{+}$ is the coordinate of the event horizon) has been constructed
for the positive-definite metric signature. Once the stress-energy tensor is calculated
it can be transformed to the physical spacetime by analytic continuation. In the exterior
region it is the familiar Wick rotation, which affects only the time coordinate.
On the other hand, inside the event horizon the problem is more complicated. 

The classical $D-$dimensional solution describing interior of the Schwarzschild-Tangherlini
black hole with the event 
horizon located at $T=r_{+}$ is given by
the line element
\begin{equation}
 ds^{2} = -\left[ \left( \frac{r_{+}}{T}\right)^{D-3}-1 \right]^{-1} dT^{2} +
 \left[\left(\frac{r_{+}}{T}\right)^{D-3}-1\right] dX^2 +  T^{2} d\Omega^{2}_{D-2},
    \label{line_classical}
\end{equation}
where $d\Omega^{2}_{D-2}$ is a metric on a unit $(D-2)$-dimensional sphere. Since only in $D=4$
case there is a simple
linear relation between the mass and $r_{+}$ in the present paper we use (almost) exclusively 
the latter. The radius of the event horizon of the Schwarzschild-Tangherlini black hole 
characterized by the mass $M$ is given by
\begin{equation}
 r_{+} = \left( \frac{16\pi G_{(D)} M}{c^{2} (D-2) \omega_{D-2}} \right)^{1/(D-3)},
\end{equation}
where $G_{(D)}$ is $D$-dimensional Newton constant and $\omega_{D-2}$ is the volume of the unit 
$(D-2)$-dimensional sphere.
If the $(D-2)$-dimensional sphere is covered by a standard ``angular'' coordinates 
$\theta_{1},...,\theta_{D-2}$
the metric $ T^{2} d\Omega^{2}_{D-2}$ can be written in the form
\begin{equation}
T^{2} d\Omega^{2}_{D-2} =  T^{2}\left[d\theta_{1}^2+ \sin^{2} \theta_{1} d\theta_{2}^{2} 
\left( ...\right)\right].
\end{equation}
Now, in order to construct the positive-definite metric let us replace $T$ by $i T,$  
$\theta_{1}$ by $i \theta_{1}$ and $r_{+}$
by $i r_{+}.$ The metric thus becomes:
\begin{equation}
  ds^{2} = \left[ \left( \frac{r_{+}}{T}\right)^{D-3}-1 \right]^{-1} dT^{2} +
 \left[\left(\frac{r_{+}}{T}\right)^{D-3}-1\right] dX^2 +  T^{2} d\Omega^{2}_{D-2},
    \label{line_classical}
\end{equation}
where 
\begin{equation}
T^{2} d\Omega^{2}_{D-2} =  T^{2}\left[d\theta_{1}^2+ \sinh^{2} \theta_{1} d\theta_{2}^{2}
\left( ...\right)\right].
\end{equation}
Note that our transformation differs form that of Ref.~\cite{Candelas1985}, which results 
in the {\it negative-}definite metric.

Having Euclidean version of the geometry of the black hole interior on can construct the one-loop
approximation to the effective action of the quantized massive fields in a large mass limit. Indeed,
for a sufficiently massive fields, i.e., when the Compton length, $\lambda_{C},$ associated 
with the mass of the field, $m,$ is much smaller 
than the characteristic radius of the curvature of the spacetime $ L,$  the contribution 
of the vacuum polarization to the effective action dominates and the contribution of real particles
is negligible. One can therefore make use of the Schwinger-DeWitt asymptotic expansion that approximates 
the effective action $W^{(1)}.$ This approach has been successfully applied 
in a number
of interesting cases and the background spacetimes range from black 
holes ~\cite{FZ1,FZ2,FZ3,JaD61,Kocio1,Matyjasek2006,LemosT,Folacci}
to cosmology~\cite{frwl2013,frwl2014} and from wormholes~\cite{Taylor} to topologically 
nontrivial spacetimes~\cite{Kofman1,Kofman2,KofmanS,MatyjasekAds}.
For the purposes of the present paper, the most relevant are the results presented in 
Refs.~\cite{HiscockInside1997,InsideJa,Matyjasek2015}

In Ref.~\cite{Matyjasek2015} it has been shown that the approximate one-loop effective action 
$W^{(1)}$ of the quantized massive scalar field
in a large mass limit can be constructed from the (asymptotic) Schwinger-DeWitt representation 
of the Green function and in the lowest order it can be written in the following form:
\begin{equation}
 W_{reg}^{(1)} =  \frac{1}{2 (4 \pi)^{D/2}} \int d^{D}x \sqrt{g} 
 \frac{[a_{k}]}{(m^{2})^{k-D/2}}\Gamma\left(k-\frac{D}{2}\right),
 \label{Wreg}
\end{equation}
where
$k= \lfloor \frac{D}{2}\rfloor +1$
and ${\lfloor} x {\rfloor}$ denote the floor function, i.e., it gives 
the largest integer less than or equal to $x.$ Here $[a_{k}]$ is the 
coincidence limit of the $k$-th Hadamard-DeWitt coefficient constructed 
from the Riemann tensor, its covariant derivatives up to $(2 k-2)$-order 
and contractions. For the technical details concerning construction of 
the Hadamard-DeWitt coefficients the reader is referred to 
Refs.~\cite{sakai,gilkey,Amsterdamski,Avramidi}. The (regularized) 
stress-energy tensor can be calculated from the standard definition 
\begin{equation}
 T^{ab} = \frac{2}{g^{1/2}} \frac{\delta}{\delta g_{ab}} W_{reg}^{(1)}.
\end{equation}

There is one-to-one correspondence between the order of the WKB approximation 
and the order of the Schwinger-DeWitt expansion.  For example, the sixth-order 
WKB approximation is equivalent to $m^{-2}$ term in $D=4$ and to $m^{-1}$ in 
$D=5$ whereas for the analogous results in $D=6$ and $D=7$ the eight-order 
WKB approximation is required. 

On general grounds one expects that the lowest-order
(nonvanishing) term of the Schwinger-DeWitt expansion is the most important.
The condition $\lambda_{C}/L \ll 1$ (with the physical constants reinserted) 
leads to
\begin{equation}
 T \gg \left( \frac{\hbar^{2} r_{+}^{D-3}s^{1/2}}{c^{D+1} m^{2}} \right)^{1/(D-1)} = 
 \left( \frac{G_{(D)} \hbar^{2} M}{ c^{D+3} m^{2}} \right)^{1/(D-1)}
 \left( \frac{16\pi  s^{1/2}}{(D-2) \omega_{D-2}} \right)^{1/(D-1)},
\end{equation}
where $s=(D-1)(D-2)^{2}(D-3)$ and $T$  is given in seconds\footnote{This 
is a generalization of the condition $T\gg (M/m^{2})^{1/3}$ employed
in the $D=4$-dimensional  back reaction calculations reported in Ref.~\cite{HiscockInside1997}}. 
For example, 
taking $D=4,$ $r_{+}$ equal to the Schwarzschild radius of the Sun and 
$m =10^{-30}$ kg one has $T\gg 10^{-16}$ which is many orders of magnitude
smaller than the coordinate time of the event horizon. It follows than 
that in our calculations we can go fairly close to the central singularity. 
Note that the coordinate time goes form $r_{+}/c$ to 0. In the rest of the 
paper we use the geometric units and the adopted conventions are those 
of Misner, Thorne and Wheeler~\cite{MTW}.

The paper is organized as follows. In Sec.~\ref{class} we study some aspects
of the classical interior of the $D$-dimensional Schwarzschild-Tangehrlini black holes.
In Sec.~\ref{zwr} we construct and formally solve the $D$-dimensional semiclassical
Einstein field equation and analyze the problem of the finite renormalization.
In Sec. \ref{gen_cons} we show how to construct the appropriate measure of anisotropy 
and investigate the two useful scalars: the Kretschmann scalar and the Karlhede scalar. 
Finally, taking the stress-energy tensor of the quantized massive scalar field, in Sec.~\ref{sec_back} 
we study the semiclassical equations and analyze the influence of quantum perturbations 
on the black hole interior for $4\leq D\leq 7.$

\section{Interior of classical Schwarzschild-Tangherlini black hole}
\label{class}
To gain a better understanding of the classical interior of the Schwarzschild-Tangherlini 
black hole let us introduce the proper time
\begin{equation} 
 \tau = \int\frac{dT}{\left[ \left( \frac{r_{+}}{T}\right)^{D-3}-1 \right]^{1/2}}
 \label{tau_plus}
\end{equation}
and, in the neighborhood of a point $\left(\theta_{(0)1}, \theta_{(0)2}, ..., \theta_{(0)D-2}\right),$
the locally Euclidean coordinates
\begin{eqnarray}
 x_{1} &=& r_{+} \left( \theta_{1} -\theta_{(0)1} \right) \nonumber \\
 x_{2} &=&  r_{+} \sin \theta_{(0)1} \left( \theta_{2} -\theta_{(0)2} \right) \nonumber \\
 &&..........................\nonumber \\ 
 x_{D-2} &=& r_{+} \sin \theta_{(0)1}...\sin \theta_{(0)D-3}  \left( \theta_{D-2} -\theta_{(0)D-2} \right).
\end{eqnarray}
Near the singularity the Schwarzschild-Tangherlini metric can be approximated by the Kasner metric
\begin{equation}
 ds^{2} = -d\tau^{2} + \left(\frac{\tau}{\tau_{0}} \right)^{-2 p_{1}} dX^{2} +  
 \left(\frac{\tau}{\tau_{0}} \right)^{2 p_{2}}\left( dx_{1}^{2} + \ldots + dx_{D-2}^{2} \right),
\label{kasner1}
 \end{equation}
where 
\begin{equation}
\tau_{0} = \frac{2 r_{+}}{D-1}, \hspace{1cm} p_{1} = -\frac{D-3}{D-1} \hspace{1cm} {\rm and} \hspace{1cm} p_{2} = \frac{2}{D-1}.
\end{equation}
It can easily be checked that both Kasner conditions are satisfied. Indeed, 
\begin{equation}
 p_{1} + (D-2) p_{2} = 1
\end{equation}
and
\begin{equation}
 p_{1}^{2} + (D-2) p_{2}^{2} =1.
\end{equation}
On the other hand, 
near the event horizon the Schwarzschild-Tangherlini metric asymptotically approaches
\begin{equation}
 ds^{2} = -d\tau^{2} + \left(\frac{D-3}{2 r_{+}} \right)^{2} \tau^{2} dX^{2} +  dx_{1}^{2} + \ldots + dx_{D-2}^{2},
 \label{kasner2}
\end{equation}
where
\begin{equation}
 \tau = \frac{2 r_{+}}{D-3} \left(1-\frac{T}{r_{+}} \right)^{1/2}.
\end{equation}
Once again it is the Kasner metric with $p_{1} =1$ and vanishing remaining Kasner exponents.
Finally observe, that the line element (\ref{kasner2}) can be formally obtained from the Rindler solution
\begin{equation}
 ds^{2} = -g x^{2} dt^{2} + dx^{2} + ...
\end{equation}
by using the complex coordinate transformation.
%%%%%%%%%%%%%%%%%%%%%%%%%%%%%%%%%%%%%%%%%%%%%%%%%%%%%%%%%%%%%%
%%%\ begin{equation}
%%%  x = i T \hspace{0.5cm} {\rm and} \hspace{0.5cm} t =X.
%%% \end{equation}
%%%%%%%%%%%%%%%%%%%%%%%%%%%%%%%%%%%%%%%%%%%%%%%%%%%%%%%%%%%%%%%

Now, let us consider two points at the same coordinate instant separated by $\Delta X.$
While the coordinate distance remains constant the physical distance between two points on the $X-$ coordinate line is given by 
\begin{eqnarray}
d_{X_{1}X_{2}} &=& \int_{X_{1}}^{X_{2}} \left[\left(\frac{r_{+}}{T}\right)^{D-3}-1 \right]^{1/2} dX\nonumber \\
&=& \left[\left(\frac{r_{+}}{T}\right)^{D-3}-1 \right]^{1/2} \Delta X
\end{eqnarray}
and grows as the coordinate time decreases.
On the other hand, the proper distance between two points separated by $d\Omega_{D-2}$ is given by
\begin{equation}
d_{\Omega} = T d\Omega_{D-2},
\end{equation}
and it decreases as the coordinate time goes from $r_{+}$ to 0. This behavior is independent of the dimension
$(D\geq 4).$

Let us return to the proper time: It should be noted that taking a positive sign of the root of the equation
\begin{equation}
d\tau^{2} = - g_{TT}\, dT^{2},
            \label{proper_time}
\end{equation}
 as it has been done in Eq.~(\ref{tau_plus}),  the proper time monotonically grows with the coordinate time $T.$
Conversely, taking the negative root, the proper time increases  as the coordinate time goes from $r_{+}$ to 0. Since the functional
relations between $\tau$ and $T$ are not very illuminating we present them graphically (Fig I), demanding 
for both types of the universe $\tau = \pi/2$ as $T=r_{+}.$ This can always be done by
 suitable choice of the integration constant. 
The universe inside the event horizon (in both time scales) has a finite lifetime. From the results collected in Table I
one sees that $\tau/r_{+}$ decreases with the dimension.
\begin{figure} %[h]
\centering					
\includegraphics[width=11cm]{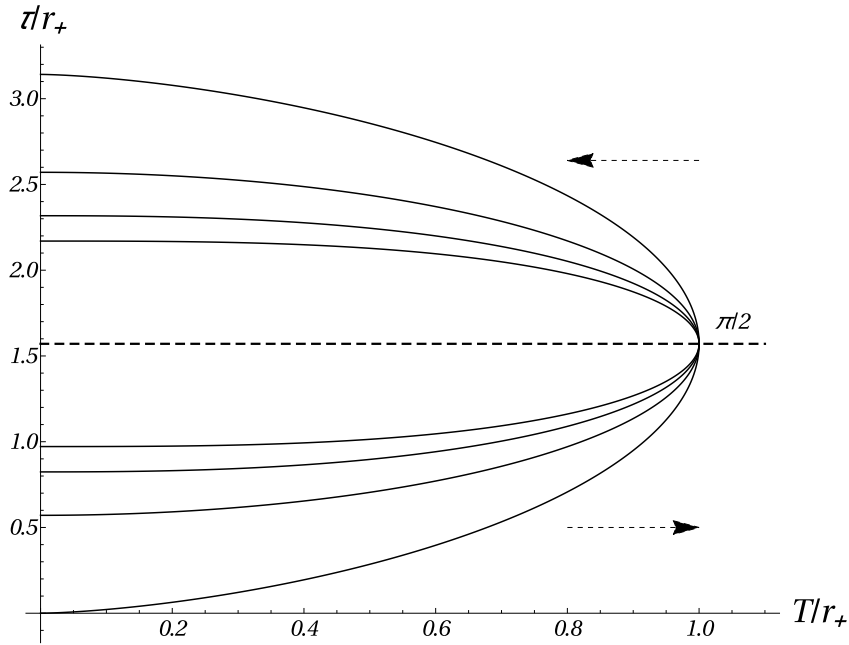}
\caption{ This graph shows the proper time $\tau$ as a function of the coordinate time $T$ for $4\leq D \leq 7.$
The upper branches are plotted for a negative root of (\ref{proper_time}) whereas the lower ones for a positive root. The arrows indicate 
the direction of the flow of the coordinate time. Top to bottom (for the upper branches) the curves are plotted for $D =4,5,6,7.$ }
\label{rys0}
\end{figure}
% \usepackage{multirow}
\begin{table}[h]
\begin{tabular}{|c|l|c|}
\hline
Dimension & Proper time           & Coordinate time               \\
\hline\hline
$D=4$         & $\tau =\pi r_{+}/2$ & \multirow{4}{*}{$T=\,r_{+}\,$} \\\cline{1-2}
$D=5$         & $\tau = r_{+}$      &                   \\\cline{1-2}
$D=6  $       & $\tau = \,0.747\, r_{+} \, $&                   \\\cline{1-2}
$D=7$         & $ \tau = 0.599 \,r_{+}$  &     \\ \hline           
\end{tabular}
\caption{The lifetime of the interior universe.}
\label{tta1}
\end{table}

%%%%%%%%%%%%%%%%%%%%%%%%%%%%%%%%%%%%%%%%%%%%%%%%%%%%%%%%%%%%%%%%%%%%%%%%%%%%%%%%%%%%%%%%%%%%%%
%\textcolor{red}{Uzupelnic o zwiazki wspolrzednych $x_{i}$ ze wspolrzednymi katowymi}
%
%\textcolor{red}{Zwiazki zewnetrznego klina Rindlera z Kasnerem w okolicach horyzontu zdarzen}
%
%\textcolor{red}{rate of deformations -- w duchu szkoly moskiewskiej: Zel'manov-Novikov-Frolov}
%
%\textcolor{red}{stosunek stalych Hubble'a}
%%%%%%%%%%%%%%%%%%%%%%%%%%%%%%%%%%%%%%%%%%%%%%%%%%%%%%%%%%%%%%%%%%%%%%%%%%%%%%%%%%%%%%%%%%%%%%
\section{The back reaction}
\label{zwr}

%%%%%%%%%%%%%%%%%%%%%%%%%%%%%%%%%%%%%%%%%%%%%%%%%%%%%%%%%%%%%%%%%%%%%%%%%%%%%%%%%%%%%%%%%%%%%
%\textcolor{red}{Umowa znakowa: $C_{1}$ jest wziete ze znakiem $+$, dokladnie tak 
%jak to ma miejsce w powyższym rownaniu}
%%%%%%%%%%%%%%%%%%%%%%%%%%%%%%%%%%%%%%%%%%%%%%%%%%%%%%%%%%%%%%%%%%%%%%%%%%%%%%%%%%%%%%%%%%%%%

The classical Schwarzschild-Tangherlini line element is a solution of the 
$D$-dimensional vacuum Einstein field equations. In this section we shall analyze the 
corrections to the characteristics of the classical black hole interior
caused by the quantum fields. The semi-classical field equations have the form
\begin{equation}
 R_{ab} -\frac{1}{2}R g_{ab} = 8 \pi T_{ab},
            \label{semi}
\end{equation}
where $T_{ab}$ is the properly regularized stress-energy tensor of the quantized field(s) and 
all remaining symbols have their usual meaning. We have chosen,
for simplicity, to work with the minimal generalization of the Einstein equations. 
Other curvature invariants can be added to the action functional, but the resulting equations
can be treated in  precisely the same way as the ``minimal'' theory.

\subsection{General considerations}
\label{gen_cons}
We shall analyze how far one can go with the semiclassical Einstein field
equations without defining explicitly the stress-energy tensor of the 
quantized fields.  The only requirement placed on the stress-energy tensor is 
its regularity on the event horizon and the absence of the net fluxes.
Unfortunately, except for metrically simple manifolds with a high degree of 
symmetry, the equations (\ref{semi}) cannot be solved exactly. However, 
assuming the expected quantum corrections to be small, one can try to solve the
equations perturbatively and concentrate on the first-order calculations 
(with the zeroth-order being the classical solution). 
To achieve this let us consider the general line element
\begin{equation}
 ds^{2}=- f(T) dT^{2} + h(T) dX^{2} + T^{2} d\Omega^{2}_{D-2}
   \label{line_semi}
\end{equation}
with
\begin{equation}
 f(T) = f_{0}(T) \left(1+ \varepsilon f_{1}(T) \right)
 \label{f_f1}
\end{equation}
and
\begin{equation}
  h(T) = h_{0}(T) \left(1+ \varepsilon h_{1}(T) \right),
  \label{h_h1}
\end{equation}
where $f_{1}(T)$ and $h_{1}(T)$ are unknown functions, and $\varepsilon$ is
a small dimensionless parameter, which helps to keep track of the order of the terms
in complicated expansions. It must not be confused (in $D=4$ case) with the small parameter
of Ref.~\cite{HiscockInside1997}. The parameter $\varepsilon$ should be set to 1 at the end of the calculations.
The functions $f_{0}(T)$ and $h_{0}(T)$ are given by $-g_{TT}$ and $g_{XX}$ of the line 
element (\ref{line_classical}), respectively.

The resulting semi-classical Einstein field equations for the line element (\ref{line_semi}-\ref{h_h1})
are given by
\begin{equation}
 \frac{d}{dT}\left[ (r_{+}^{D-3} - T^{D-3} ) f_{1}(T) \right] = \frac{16 \pi}{D-2} T^{D-2} T_{X}^{X} 
 \label{row1}
\end{equation}
and
\begin{equation}
 \frac{d}{dT} h_{1}(T) =-\frac{(D-3) T^{D-4}}{r_{+}^{D-3} - T^{D-3}} f_{1}(T) 
 -\frac{16\pi}{D-2} \frac{T^{D-2}} { r_{+}^{D-3}-T^{D-3}} T_{T}^{T}.
 \label{row2}
\end{equation}
The first equation can formally be integrated to give
\begin{equation}
 f_{1}(T) = \frac{C_{1}}{r_{+}^{D-3}-T^{D-3}} + \frac{16 \pi }{(D-2)(r_{+}^{D-3}-T^{D-3})} \int_{r_{+}}^{T} T^{D-2} T_{X}^{X} dT .
         \label{fun_f1}
 \end{equation}
It can easily be shown that the integration constant $C_{1}$ has no independent 
meaning and can be absorbed into the definition of the renormalized (dressed) radius 
of the event horizon. Moreover, by the very same procedure, the constant $C_{1}$ can be absorbed 
in the second equation. Let us analyze this problem more closely. First, consider the function $f(T),$ which can be written as
\begin{equation}
 1/f(T) = -\left[  \left(  \frac{r_{+}}{T} \right)^{D-3} -1 -\frac{\varepsilon C_{1}}{T^{D-3}} -
 \frac{16\pi \varepsilon}{(D-2) T^{D-3}}  \int_{r_{+}}^{T} T^{D-2} T_{X}^{X} dT  \right]
 + {\cal O}(\varepsilon^{2})
\end{equation}
and observe that introducing the renormalized radius of the event horizon,
 $\bar{r}_{+},$ defined by the equation
\begin{equation}
 r_{+} \to \bar{r}_{+} = r_{+} + \frac{\varepsilon C_{1}}{(D-3) r_{+}^{D-4}}
\end{equation}
the integration constant can be relegated in the first-order calculations.
The same transformation can be used to renormalize $r_{+}$ in the second metric potential 
\begin{equation}
 h(T) = \left( \frac{r_{+}}{T} \right)^{D-3}  -\frac{\varepsilon C_{1}}{T^{D-3}} -1 
+ \varepsilon C_{2} +  {\cal O}(\varepsilon),
\end{equation}
where ${\cal O}(\varepsilon)$ terms containing integrals of 
the stress-energy tensor have not been displayed explicitly. 
To determine the second integration constant, $C_{2},$ additional 
piece of information is needed. Fortunately,
considered characteristics of the quantum-corrected interior of 
the Schwarzschild-Tangherlini black holes are independent of $C_{2}.$
Since $C_{1}$ and $r_{+}$ have no independent physical meaning, in what follows, 
for notational simplicity, we shall replace $\bar{r}_{+}$ with $r_{+}$ and treat
$r_{+}$ as the renormalized (dressed) radius of the event horizon.

%%%%%%%%%%%%%%%%%%%%%%%%%%%%%%%%%%
%%%%%%%%%%%%%%%%%%%%%%%%%%%%%%%%%
On general grounds one expects that the components of the stress-energy tensor 
constructed within the framework of the Schwinger-DeWitt approximation are simple 
polynomials in $r_{+}/T,$ and hence the calculations of the functions $f_{1}$ and $h_{1}$
reduce to two elementary quadratures. Now, in order to better understand the influence 
of the quantized fields on the black hole interior, we shall study the trace 
of the rate of the deformations tensor and 
the ratio of the Hubble parameters. Similarly, to study the influence of the quantized fields
on the curvature  we calculate  the Kretschmann scalar. Additionally we will check if the
Karlhede's scalar is still a useful device for detecting the event horizon.
 
The interior of the Schwarzschild-Tangherlini black holes is nonstatic and anisotropic. 
Following Novikov's  paper~\cite{novikov1961} this can be analyzed using 
the rate of deformations tensor. Let us introduce the tensor $p_{ab}$ defined as
\begin{equation}
p_{ab} = g_{ab} + u_{a} u_{b},
\end{equation}
where $u^{a} = (-g_{00})^{-1/2}\delta^{a}_{0}.$
Let the indices from the second half of the Latin alphabet denote spatial coordinates. 
The deformation rate tensor, which has only spatial components,  is given by~\cite{nov-frol}
\begin{equation}
{\cal D}_{rs} = \frac{1}{2 \sqrt{-g_{00}}} \frac{\partial}{\partial T}\,p_{rs}
\end{equation}
and its trace is ${\cal D} = {\cal D}_{r}^{r} .$
Now, let us consider a volume element
$
vol = \sqrt{p} \,\Delta X \,\Delta \theta_{1} ...\Delta\theta_{D-2},
$
where $p = det(p_{rs}),$ and construct the quantity
\begin{equation}
\sqrt{-g_{00}}\,{\cal D} = \frac{1}{vol}  \,\frac{\partial}{\partial T} vol
\end{equation}
with a natural interpretation as the speed of the relative change of the volume element of the space.
For the quantum-corrected Schwarzschild-Tangherlini black hole  the trace of the rate of deformation 
tensor ${\cal D}$ is given by 
 \begin{equation}
  {\cal D} = {\cal D}_{0} + \varepsilon {\cal D}_{1},
 \end{equation}
where
\begin{equation}
  {\cal D}_{0} =\frac{1}{T}\left[ \left( \frac{r_{+}}{T}\right)^{D-3}-1 \right]^{-1/2} 
\left[\left(\frac{D-1}{2}\right)\left(\frac{r_{+}}{T}\right)^{D-3}-(D-2)\right]
\end{equation}
and
\begin{equation}
  {\cal D}_{1} = - \frac{1}{2} f_{1}(T) {\cal D}_{0}  + \frac{1}{2 }
 \left[ \left( \frac{r_{+}}{T}\right)^{D-3}-1 \right]^{1/2} h'_{1}(T).
\end{equation}
Te trace ${\cal D}$ is independent of the integration constant $C_{2}.$ It should be noted that in the closest vicinity 
of the event horizon the correction to the trace ${\cal D}$ is practically independent of the function $h'_{1}.$ 
On the other hand, the behavior of the correction ${\cal D}_{1},$ i.e., the question if the rate of deformations 
grows or decreases depends on the sign of the $f_{1}.$ 

The conclusions that can be drawn from the analysis of the classical part of the tensor ${\cal D}_{rs}$ and its 
trace are qualitatively similar to those of the Novikov's paper.\footnote{It should be noted that English
translation of the Novikov's paper is not always correct. } Regardless of the dimension ${\cal D}_{0}$ is 
always negative in the vicinity of the event horizon, whereas it is always positive for $T<T_{min} \approx 0.75 r_{+}.$
$T_{min}$ is the smallest root of ${\cal D}_{0}(T) =0.$

 Although the information yielded by the rate of deformation
tensor is accurate, it is simultaneously hard to visualize and we need something somewhat simpler.
The useful measure of the anisotropy is the ratio of the Hubble parameters
\begin{equation}
 \alpha = \frac{H_{X}}{H_{\theta}} = \frac{g_{\theta \theta} \frac{d }{dT}g_{XX} }{g_{XX} \frac{d }{dT} g_{\theta \theta}},
\end{equation}
where $\theta$ is any of the angular coordinates
Making use of  (\ref{f_f1}) and (\ref{h_h1}), one obtains 
\begin{equation}
 \alpha = \frac{D-3}{2} \frac{\left(\frac{r_{+}}{T}\right)^{D-3}}
 {1- \left(\frac{r_{+}}{T}\right)^{D-3}} + \frac{1}{2}T h_{1}'(T),
\end{equation}
where the first term in the right hand side of the above equation is the unperturbed part of 
$\alpha.$ Consequently, the second term, which we denote by $\delta \alpha$, depending on the sign can make the black hole interior
more isotropic or anisotropic. Further analysis of the role played by $\delta \alpha$ 
must be postponed until we solve the semicalssical Einstein field equations.

It could be easily shown that the simple differential curvature invariant 
which is very useful in detecting the location of $r_{+}$ (sometimes called Karlhede's invariant~\cite{Karlhede})
\begin{equation}
 I =R_{abcd;e} R^{abcd;e}
\end{equation}
vanishes on the event horizon of the Schwarzschild-Tangherlini black hole and is positive
inside and negative outside. 
Because of their properties such invariants have become  popular recently, see e.g.,~\cite{PageShoom,Lake,Moffat}.
Now, making use of the functions $f$ and $h$ one has
\begin{eqnarray}
 I &=& (D-3)(D-2)(D-1) \left(\frac{r_{+}}{T}\right)^{D-3}\left[\left( \frac{r_{+}}{T}\right)^{D-3}-1\right]\left\{  (D-2)(1-D^{2}) \left( \frac{r_{+}}{T}\right)^{D-3} \frac{1}{T^{6}}\right.\nonumber \\
 && + \left[10 + (-4 + 7 D - 3 D^{2}) \left( \frac{r_{+}}{T}\right)^{D-3} \right] \frac{ f'_{1}(T)}{T^{5}}- \left[ 6 + D(7 - 3 D) \left( \frac{r_{+}}{T}\right)^{D-3} \right] \frac{h'_{1}(T)}{T^{5}}\nonumber \\
 && -\left[2 + (1 - D) \left( \frac{r_{+}}{T}\right)^{D-3} \right] \frac{f''_{1}(T)}{T^{4}} + \left[ 6 + (9 - 5 D) \left( \frac{r_{+}}{T}\right)^{D-3} \right] \frac{h''_{1}(T)}{ T^{4}} \nonumber \\
 &&  \left. +  2 \left[  \left( \frac{r_{+}}{T}\right)^{D-3}-1 \right] \frac{h_{1}^{(3)}(T)}{ T^{3}} +  \left[3 (2 - D - 2 D^{2} + D^{3}) \left( \frac{r_{+}}{T}\right)^{D-3}-16  \right] \frac{f_{1}(T)}{T^{6}}\right\}.
\label{Iniv}
 \end{eqnarray}
To obtain the classical term it suffices to set to zero the functions $f_{1}(T)$ and $h_{1}(T).$
Because of the  presence of the factor  $h_{0}(T),$ the invariant $I$ of the quantum-corrected 
black hole always vanishes at the event horizon provided the functions $f_{1}(T)$ and $h_{1}(T)$ 
are regular in its vicinity. In view of the foregoing discussion it is  expected that 
in the case at hand the condition $I(r_{+}) =0$ is satisfied.

Finally,  let us consider the simplest curvature invariant, namely the Kretschmann scalar,
defined as the ``square'' of the Riemann tensor
\begin{equation}
 K = R_{abcd}R^{abcd},
\end{equation}
which, for the quantum-corrected interior of the $D$-dimensional Schwarzschild-Tangherlini black hole  
has the following form:
\begin{eqnarray}
 K &=& \frac{ (D-1)(D-2)^{2}(D-3)}{r_{+}^{4}}\left( \frac{r_{+}}{T}\right)^{2D-2} 
 +\frac{2(D-2)(D-3)}{r_{+}^{2}} \left[ \left(\frac{r_{+}}{T} \right)^{2 D-4} 
 -\left( \frac{r_{+}}{T}\right)^{D-1} \right] h_{1}''(T) 
 \nonumber \\
 && - \frac{2(D-2)(D-3)}{r^{3}_{+} T}\left( \frac{r_{+}}{T}\right)^{D}\left[
 -2 + 2 \left( \frac{r_{+}}{T}\right)^{D-3} -3 D \left( \frac{r_{+}}{T}
 \right)^{D-3} +D^{2} \left( \frac{r_{+}}{T}\right)^{D-3}\right] f_{1}(T) \nonumber \\
 &&+ \frac{(D-2)(D-3)}{r_{+}^{3}} \left( \frac{r_{+}}{T}\right)^{D}\left[ 
 -2 - \left(\frac{r_{+}}{T}\right)^{D-3} + D  \left(\frac{r_{+}}{T}\right)^{D-3} \right] f_{1}'(T)
 \nonumber \\
 &&-\frac{(D-2)(D-3)}{r_{+}^{3}} \left( \frac{r_{+}}{T}\right)^{D}\left[ -2 - 7 
 \left(\frac{r_{+}}{T}\right)^{D-3} + 3D  \left(\frac{r_{+}}{T}\right)^{D-3} \right] h_{1}'(T).
\end{eqnarray}
The first term in the right-hand-side of the above equation gives the classical Kretschmann 
scalar and the remaining ones are the quantum corrections, which we denote by $\delta K.$ Although the semiclassical Einstein 
field equation are certainly incorrect as $T \to 0,$ and should be replaced by the (unknown as yet)
quantum gravity, it is of some interest to study the tendency exhibited by $\delta K$ in this very limit.
This, however, requires explicit knowledge of the functions $f_{1}(T)$ and $h_{1}(T),$ which
is the subject of the next subsection.

\subsection{The back reaction of the quantized massive fields}
\label{sec_back}

Now, let us return to the semiclassical Einstein field equations and solve 
(\ref{row1}) and (\ref{row2}) with the stress-energy tensor of the quantized  
massive scalar field. The relevant components of $T_{a}^{b}$ are listed in Appendix~\ref{appA}.
The angular components can easily be calculated form the covariant conservation 
equation $\nabla_{a} T^{a}_{b} =0.$
For any considered dimension, the components of the tensor  are simple
polynomials  (in $x = r_{+}/T$),   the difference   $T^{T}_{T} -T^{X}_{X}$ 
factorizes as 
\begin{equation}
 T^{T}_{T} -T^{X}_{X} = F(T) g_{XX},
\end{equation} 
and the function $F(T)$ is regular for $T>0.$  
Indeed, after some algebra, one has
\begin{equation} 
 g_{00} = -\left( x^{D-3}-1\right)^{-1} \left(1 + \varepsilon f_{1} \right)
\end{equation}
and
\begin{equation}
 g_{11} = \left( x^{D-3} -1\right) \left( 1+ \varepsilon h_{1}\right),
\end{equation}
where for $D=4$ 
\begin{equation}
 f_{1} = \alpha^{(4)}_{1} x (1+x +x^{2} +x^{3} +x^{4}) +\beta^{(4)}_{1} x^{6},
 \label{f1}
\end{equation}
\begin{equation}
 h_{1} = -\alpha_{1}^{(4)} x (1+ x+x^{2}+x^{3}+x^{4}) + \beta^{(4)}_{2} x^6 + K^{(4)},
\end{equation}
 and
where the  coefficients $\alpha_{i}^{(D)}$ and the  integration constants $K^{(D)}$ are 
listed in Appendix~\ref{appB}. This result is not new: The stress-energy tensor 
has been obtained in the early 80's by Frolov and Zel'nikov~\cite{FZ1,FZ2} and  subsequently 
used in Ref.~\cite{HiscockInside1997}. 
To the best of our knowledge the results for higher dimensional geometries are new. 
Now, making use of the stress-energy tensor constructed in the $D=5$ Schwarzschild-Tangherlini 
spacetime, one has
\begin{equation}
 f_{1} = \alpha_{1}^{(5)} x^{2} (1+x^{2} +x^{4} ) + \beta^{(5)}_{1} x^{8},
\end{equation}
\begin{equation}
 h_{1} = - \alpha_{1}^{(5)} x^{2} (1+x^{2} +x^{4} ) + \beta^{(5)}_{2} x^{8} + K^{(5)}.
\end{equation}
Both tensors have been calculated from the effective action constructed from the $[a_{3}].$
Similarly, making use the effective action constructed form the coefficient $[a_{4}],$ one obtains 
\begin{equation}
 f_{1} = \alpha_{1}^{(6)} x^{3} (1+ x^{3} +x^{6}) + \beta^{(6)}_{1} x^{12} + \gamma_{1}^{(6)} x^{15},
\end{equation}
\begin{equation}
 h_{1} = -\alpha_{1}^{(6)} x^{3} (1+ x^{3} +x^{6}) + \beta^{(6)}_{2} x^{12} + \gamma_{2}^{(6)} x^{15} + K^{(6)}
\end{equation}
and
\begin{equation}
 f_{1} = \alpha_{1}^{(7)} \left[ \frac{x^{4}}{1+x^{2}} + x^{6} (1+x^{4})\right] + \beta_{1}^{(7)} x^{14} + \gamma_{1}^{(7)} x^{18},
\end{equation}
\begin{equation}
 h_{1} =- \alpha_{1}^{(7)} \left[ \frac{x^{4}}{1+x^{2}} + x^{6} (1+x^{4})\right] + \beta_{2}^{(7)} x^{14} + \gamma_{2}^{(7)} x^{18} + K^{(7)},
\label{hlast}
 \end{equation}
for $D=6$ and $D=7,$ respectively.
It should be noted that for $D=7$ the functions loose their polynomial character, but they are still regular except for $T=0.$

\begin{figure} [h]
\centering
\includegraphics[width=11cm]{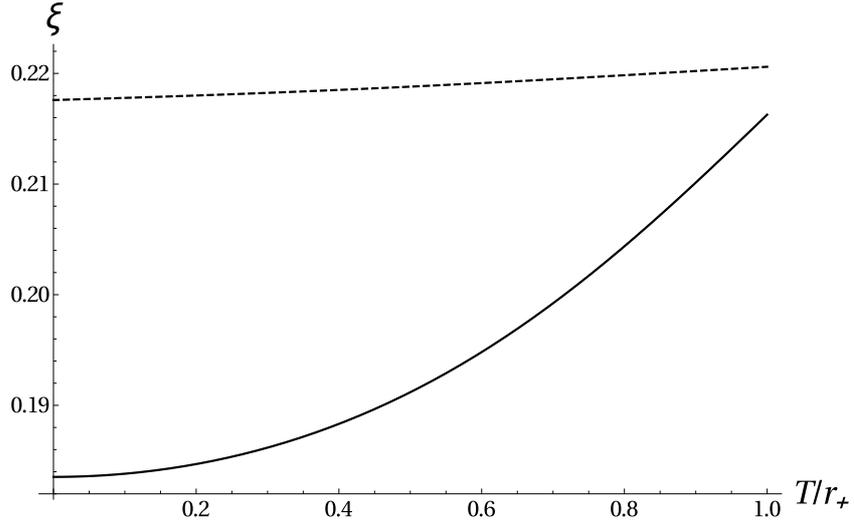}
\caption{ The dependence of the curvature coupling parameter, $\xi,$ on  $T/r_{+}$ for zero perturbation to the anisotropy $(\delta \alpha =0)$
of the interior of the $D=4$ (dashed line) and $D=5$ (solid line) Schwarzschild-Tangherlini black hole. Above the curves, the quantum corrections make the black hole
interior more anisotropic, whereas below the curves the spacetime becomes more isotropic.}
\label{rys1}
\end{figure}

\begin{figure} [h]
\centering
\includegraphics[width=11cm]{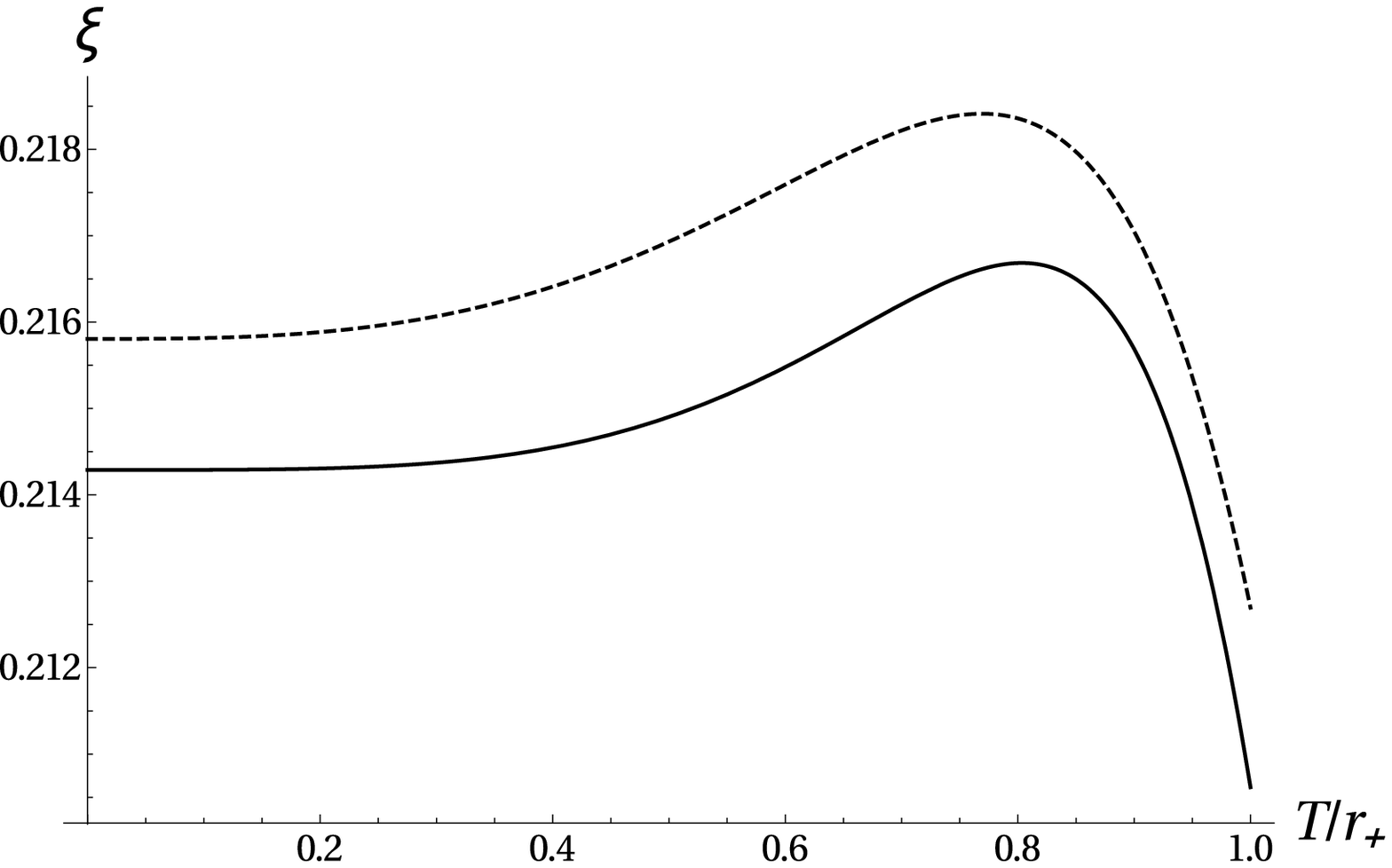}
 \caption{ The dependence of the curvature coupling parameter, $\xi,$ on  $T/r_{+}$ for zero perturbation to the anisotropy $(\delta \alpha =0)$
of the interior of the $D=6$ (dashed line) and $D=7$ (solid line) Schwarzschild-Tangherlini black hole. Above the curves, the quantum corrections make the black hole
interior more isotropic, whereas below the curves the spacetime becomes more anisotropic.} 
\label{rys2}
\end{figure}

Having constructed the functions $f_{1}(T)$ and $h_{1}(T)$ one can analyze the quantum corrections to the
Kretschmann scalar and anisotropy of the black hole interior.  First, let us consider $\alpha.$
Inspection of the unperturbed part of  $\alpha$ shows that it is always negative. The sign of $\alpha$
is positive if the internal geometry expands or contracts in all spatial directions. Of course,
for isotropic evolution one has $\alpha =1.$ The negative sign of $\alpha$ means that it is contracting in one dimension and expanding 
in the other.
Depending on the sign, the quantum  perturbation can strengthen or weaken the anisotropy. 
Since $\alpha_{0} <0$ the anisotropy is weaken for $\delta\alpha >0$ and strengthen in the regions
where $\delta\alpha <0.$ This, however, depends on the coupling constant $\xi$ and the coordinate time $T$ in a quite complicated way. 
The results of the numerical calculations has been plotted in Figs~\ref{rys1} and \ref{rys2}. Specifically, for $D=4$ and $D=5$ the 
$\delta\alpha$ is negative above the $(\delta\alpha=0)$-curves and positive below. On the other hand, for $D=6$ and $D=7$ 
the perturbation is negative below the curves and positive above. 

\begin{table}[h]
\begin{tabular}{|c|c|c|}
\hline
 Dimension   & $\xi =0$                           & $\xi =\frac{D-2}{4D-4}$      \\ \hline\hline
$D=4$ & more isotropic   & more isotropic               \\ \hline
$D=5$ & more isotropic                   & more isotropic  for  $x>0.365$             \\ \hline
$D=6$ & more anisotropic                   & more anisotropic  \\ \hline
$D=7$ & more anisotropic                   & more anisotropic  \\  \hline                          
\end{tabular}
\caption{The influence of  $\delta \alpha$ on the black hole interior. Depending on the coupling
$\xi$ and the dimension the quantum corrections can make the spacetime more isotropic or more anisotropic.}
\label{tta2}
\end{table}

Now, we shall analyze in some details the behavior of the corrections to the Kretschmann scalar and the measure
of the anisotropy $\alpha$ for the physical values of the coupling parameter, i.e., for the minimal coupling 
$\xi = 0$ and the conformal coupling $\xi_{c} = (D-2)/(4 D-4).$ 
(There is no need to perform such analysis for Karlhede's scalar as its main role is to serve as a useful device 
for detecting location of the event horizon. Inspection of (\ref{Iniv}) and (\ref{f1}-\ref{hlast}) shows that $I=0$ at the quantum-corrected event horizon, as expected.) 

As have been observed earlier in Ref.~\cite{HiscockInside1997}, the quantum corrections for the minimal and conformal coupling always 
tend to isotropize the interior of the  Schwarzschild black hole. On the other hand however, in higher dimension the pattern is more complicated.
Indeed, for $D=5$ the perturbation $\delta \alpha >0 $ for $\xi =0$ whereas for $\xi_{c} = 3/16$ isotropization occurs only for $T > 0.365 r_{+} .$
In turn, for $D=6$ and $D=7,$ the perturbation $\delta \alpha$ is always negative for the minimal and the conformal coupling, i.e.,
the quantum effects make the black hole interior more anisotropic. These results are tabulated in Table~\ref{tta2}.

\begin{table}[h]
\begin{tabular}{|c|c|c|}
\hline
 Dimension   & $\xi =0$                           & $\xi =\frac{D-2}{4D-4}$      \\ \hline\hline
$D=4$ & $\delta K > 0$ for $x>0.986 $  & $\delta K$ always positive         \\ \hline
$D=5$ & $\delta K$ always positive                  & $\delta K$ always positive            \\ \hline
$D=6$ & $\delta K$ always positive                   &  $\delta K >0$ for $x>0.543$  \\ \hline
$D=7$ & $\delta K$ always positive                 & $\delta K >0$ for $x>0.599$  \\  \hline                          
\end{tabular}
\caption{The sign of the quantum corrections to the Kretschmann scalar.}
\label{tta3}
\end{table}
 
\begin{figure} %[h]
\centering
\includegraphics[width=11cm]{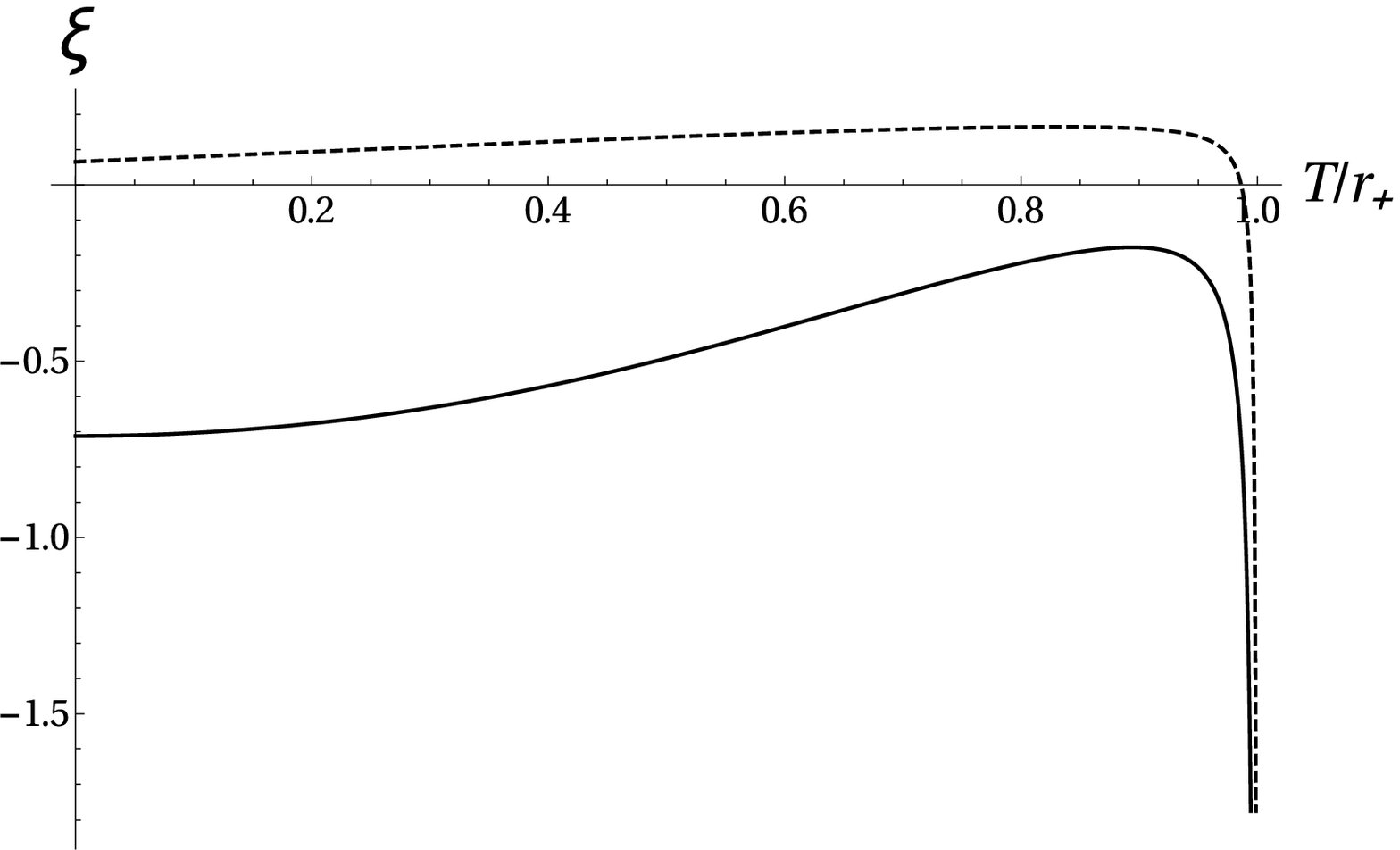}
\caption{The dependence of the curvature coupling parameter, $\xi,$ on  $T/r_{+}$ for zero perturbation to the Kretschmann scalar
of the interior of the $D=4$ (dashed line) and $D=5$ (solid line) Schwarzschild-Tangherlini black hole. Above the curves, the quantum corrections to the scalar $K$ 
are positive, whereas below the curves they are negative.  }
\label{rys3}
\end{figure}

\begin{figure} %[h]
\centering
\includegraphics[width=11cm]{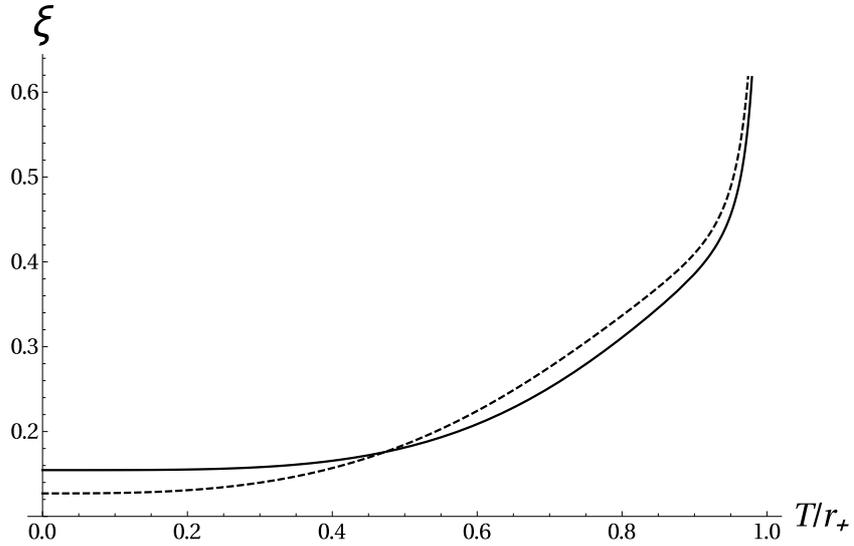}
\caption{The dependence of the curvature coupling parameter, $\xi,$ on  $T/r_{+}$ for zero perturbation to the Kretschmann scalar
of the interior of the $D=6$ (dashed line) and $D=7$ (solid line) Schwarzschild-Tangherlini black hole. Above the curves, the quantum corrections to the 
scalar $K$ are negative, whereas below the curves they are positive.   }
\label{rys4}
\end{figure}

 Let us analyze how the growth of the curvature are affected by the quantum processes. And since the classical part 
 of the  Kretschmann scalar is positive, one concludes
 that the growth of curvature (as $T$ decreases) is weakened if the perturbation is negative. Inspection of the
 Figs.~\ref{rys3} and \ref{rys4} as well as the Table~\ref{tta3} shows that initially, regardless of the dimension,  
 $\delta K$ is always positive for the both types of couplings. This is very important message as it refers to the region where the 
quality of  the approximation is expected to be high.
 
 A few words of comment are in order here. First, it should be emphasized once more that although we have plotted functions $\delta \alpha=0$
 and $\delta K =0$ for all allowable values of the coordinate time the approximation certainly does not work in the region close
 to the central singularity. Therefore our results show the tendency in behavior of the quantum corrections as the central 
 singularity is approached (which can be misleading) rather than their actual run. Of course the answer to the question of how close the singularity
 can be approached depends on many factors, such as dimension, the `radius' of the event horizon and  the  type of the quantized field. Each case should be 
 analyzed separately.
 The second observation is less obvious and is, roughly speaking, related to the question of how long the first-order approximation
 dominates the higher-orders terms inside the event horizon.  Once again this problem goes to the very core of the Schwinger-DeWitt asymptotic expansion. 
 And once again there is no better answer than to recall its  principal assumptions.  Finally, let us observe that although the quantum corrections caused 
 by a solitary field is expected to be small in the  domain of applicability of the approximation, they can be made arbitrarily large for a large
 number of fields.

%\clearpage

\clearpage

\appendix
\section{The stress-energy tensor of the quantized massive scalar field in the interior of the higher-dimensonal Schwarzschild-Tangherlini black hole }
\label{appA}

In this appendix we list the $(T,T)$ and $(X,X)$ components of the  stress-energy tensor for $4 \leq D \leq 7.$
The angular components are not displayed  as they can easily be calculated from the covariant conservation equation.

$D =4.$
\begin{equation}
 T_{T}^{(4)T} = \xi \frac{ r_{+}^{2} (3 r_{+}-4 T)}{80 \pi^2 m^2
   T^9}+\frac{r_{+}^{2} (63 T-47 r_{+})}{5760 \pi^2 m^2 T^9}
\end{equation}

\begin{equation}
 T_{X}^{(4)X} = \frac{r_{+}^2 (1237 r_{+}-1125 T)}{40320 \pi ^2 m^2 T^9}-\frac{\xi 
   r_{+}^2 (11 r_{+}-10 T)}{80 \pi ^2 m^2 T^9}
\end{equation}

%\begin{equation}
% T_{\theta}^{(4)\theta} = \frac{r_{+}^2 (1543 r_{+}-1323 T)}{40320 \pi ^2 m^2 T^9}-\frac{\xi 
%   r_{+}^2 (7 r_{+}-6 T)}{40 \pi ^2 m^2 T^9}
% \end{equation}

$D=5.$
\begin{equation}
 T_{T}^{(5)T} = \frac{\xi  r_{+}^4 \left(2 r_{+}^2-3 T^2\right)}{10 \pi ^2 m
   T^{12}}+\frac{r_{+}^4 \left(297 T^2-185 r_{+}^2\right)}{5040 \pi
   ^2 m T^{12}}
\end{equation}

\begin{equation}
 T_{X}^{(5)X} = \frac{841 r_{+}^6-729 r_{+}^4 T^2}{5040 \pi ^2 m T^{12}}-\frac{\xi 
   r_{+}^4 \left(7 r_{+}^2-6 T^2\right)}{10 \pi ^2 m T^{12}}
\end{equation}

%\begin{equation}
% T_{\theta}^{(5)\theta} = \frac{299 r_{+}^6-231 r_{+}^4 T^2}{1680 \pi ^2 m T^{12}}-\frac{\xi 
%   r_{+}^4 \left(9 r_{+}^2-7 T^2\right)}{10 \pi ^2 m T^{12}}
%\end{equation}

$D=6.$
\begin{equation}
 T_{T}^{(6)T} = \frac{r_{+}^6 \left(53938 r_{+}^6-115360 r_{+}^3 T^3+48195
   T^6\right)}{20160 \pi ^3 m^2 T^{20}}-\frac{5 \xi  r_{+}^6 \left(6665
   r_{+}^6-14444 r_{+}^3 T^3+6048 T^6\right)}{2688 \pi ^3 m^2
   T^{20}}
\end{equation}
\begin{equation}
T_{X}^{(6)X} = \frac{5 \xi  r_{+}^6 \left(11997 r_{+}^6-15956 r_{+}^3 T^3+4536
   T^6\right)}{896 \pi ^3 m^2 T^{20}}-\frac{r_{+}^6 \left(295892
   r_{+}^6-404570 r_{+}^3 T^3+121905 T^6\right)}{20160 \pi ^3 m^2
   T^{20}}
\end{equation}

%\begin{equation}
% T_{\theta}^{(6)\theta} = \frac{5 \xi  r_{+}^6 \left(21328 r_{+}^6-27157 r_{+}^3 T^3+7560
%   T^6\right)}{1344 \pi ^3 m^2 T^{20}}-\frac{r_{+}^6 \left(1387753
%   r_{+}^6-1754830 r_{+}^3 T^3+481950 T^6\right)}{80640 \pi ^3 m^2
%   T^{20}}
%\end{equation}

$D=7.$
\begin{equation}
 T_{T}^{(7)T} = \frac{r_{+}^8 \left(30549 r_{+}^8-66088 r_{+}^4 T^4+26544
   T^8\right)}{4480 \pi ^3 m T^{24}}-\frac{9 \xi  r_{+}^8 \left(198
   r_{+}^8-435 r_{+}^4 T^4+175 T^8\right)}{56 \pi ^3 m T^{24}}
\end{equation}

\begin{equation}
 T_{X}^{(7)X} = \frac{9 \xi  r_{+}^8 \left(528 r_{+}^8-672 r_{+}^4 T^4+175
   T^8\right)}{28 \pi ^3 m T^{24}}-\frac{r_{+}^8 \left(4713
   r_{+}^8-6192 r_{+}^4 T^4+1736 T^8\right)}{128 \pi ^3 m T^{24}}
\end{equation}

%\begin{equation}
% T_{\theta}^{(7)\theta} = \frac{9 \xi  r_{+}^8 \left(1254 r_{+}^8-1515 r_{+}^4 T^4+385
%   T^8\right)}{56 \pi ^3 m T^{24}}-\frac{r_{+}^8 \left(971439
%   r_{+}^8-1165928 r_{+}^4 T^4+291984 T^8\right)}{22400 \pi ^3 m
%   T^{24}}
%\end{equation}

\section{Coefficients of the functions $f_{1}(T)$ and $h_{1}(T)$}
\label{appB}
Here we list the coefficients of the functions $f_{1}(T)$ and $h_{1}(T).$
(See Eqs.~(\ref{f1} -\ref{hlast})). The integration constants $C_{2}^{(D)}$ 
are left unspecified and should be determined from the physically
motivated boundary conditions.  All quantities considered in the main text, 
such as $\alpha,$ ${\cal D},$ $K$ and $I$  are independent of the integration 
constant $C_{2}^{(D)}.$
\begin{equation}
 \alpha_{1}^{(4)} = \frac{113-504\xi}{30240\pi m^{2} r_{+}^{4}}
\end{equation}
\begin{equation}
 \beta_{1}^{(4)} = \frac{-1237+5544\xi}{30240 \pi m^{2} r_{+}^{4}}
\end{equation}
\begin{equation}
 \beta_{2}^{(4)} =-\frac{ 47-216 \xi}{4320 \pi m^{2} r_{+}^{4}}
\end{equation}
\begin{equation}
 K^{(4)} = \frac{149-672 \xi}{5040 \pi m^{2} r_{+}^{4}} +C_{2}^{(4)}
\end{equation}
\begin{equation}
 \alpha_{1}^{(5)} = \frac{131-504\xi}{7560 m \pi r_{+}^{4}}
\end{equation}
\begin{equation}
 \beta_{1}^{(5)} = - \frac{841-3528 \xi}{7560 m \pi r_{+}^{4}}
\end{equation}
\begin{equation}
 \beta_{2}^{(5)} = - \frac{185-1008\xi }{7560 m \pi r_{+}^{4}} 
\end{equation}
\begin{equation}
 K^{(5)} = \frac{289-1260 \xi}{3780 m \pi r_{+}^{4}} +C_{2}^{(5)}
\end{equation}
\begin{equation}
 \alpha_{1}^{(6)} = - \frac{13291 - 87300 \xi}{151200 m^{2} \pi^{2} r_{+}^{6}}
\end{equation}
\begin{equation}
 \beta_{1}^{(6)} = -\frac{419641 - 1788300 \xi}{151200 m^{2} \pi^{2} r_{+}^{6}}
\end{equation}
\begin{equation}
 \gamma_{1}^{(6)} = \frac{591784 - 2699325 \xi}{151200 m^{2} \pi^{2} r_{+}^{6}}
\end{equation}
\begin{equation}
 \beta_{2}^{(6)} = -\frac{5609 - 54450 \xi}{151200 m^{2} \pi^{2} r_{+}^{6}}
\end{equation}
\begin{equation}
 \gamma_{2}^{(6)} = \frac{107876 - 499875 \xi}{ 151200 m^{2} \pi^{2} r_{+}^{6}}
\end{equation}
\begin{equation}
 K^{(6)} = -\frac{9476 - 47155 \xi }{ 10080 m^{2} \pi^{2} r_{+}^{6}} + C_{2}^{(6)}
\end{equation}
\begin{equation}
 \alpha_{1}^{(7)} = -\frac{1439-10080 \xi}{8400 m \pi^{2} r_{+}^{6}}
\end{equation}
\begin{equation}
 \beta_{1}^{(7)} = -\frac{7579 -32256 \xi}{1680 m \pi^{2} r_{+}^{6}}
\end{equation}
\begin{equation}
\gamma_{1}^{(7)} =  \frac{10997 - 50688 \xi}{1680 m \pi^{2} r_{+}^{6}}
\end{equation}
\begin{equation}
 \beta_{2}^{(7)} = \frac{479 + 720 \xi}{8400 m \pi^{2} r_{+}^{6}}
\end{equation}
\begin{equation}
 \gamma_{2}^{(7)} =  \frac{10183 - 47520 \xi}{8400 m \pi^{2} r_{+}^{6}}
\end{equation}
\begin{equation}
 K^{(7)} = - \frac{28519 - 144000 \xi}{16800 m \pi^{2} r_{+}^{6}} + C_{2}^{(7)}
\end{equation}

%\bibliography{inside}

\end{document}